\documentclass[footinbib,twocolumn,showpacs,amsmath,amstex,amssymb,mathfonts,superscriptaddress,prl]{revtex4}
\usepackage{graphicx}
\usepackage{color}
\usepackage{bm}

\usepackage{graphicx}
\usepackage{color}
\usepackage{bm}
\usepackage{amsmath}
\usepackage{amssymb}
\usepackage{amsthm}
\usepackage{amsfonts}
\usepackage{multirow}
\usepackage{makecell}
\usepackage{float}
\usepackage{comment}
\usepackage{enumitem}
\usepackage{verbatim}
\usepackage{mathtools}

\usepackage{bbm}

\begin{document}

\title{Statistical interactions and boson-anyon duality in fractional quantum Hall fluids}
\author{Bo Yang} 
\affiliation{Division of Physics and Applied Physics, Nanyang Technological University, Singapore 637371.}
\affiliation{Institute of High Performance Computing, A*STAR, Singapore, 138632.}

\date{\today}
\begin{abstract}
We present an exact scheme of bosonization for anyons (including fermions) in the two-dimensional manifold of the quantum Hall fluid. This gives every fractional quantum Hall phase of the electrons one or more dual bosonic descriptions. For interacting electrons, the statistical transmutation from anyons to bosons allows us to explicitly derive the microscopic statistical interaction between the anyons, in the form of the effective two-body and few-body interactions. This also leads to a number of unexpected topological phases of the single component bosonic fractional quantum Hall effect that may be experimentally accessible. Numerical analysis of the energy spectrum and ground state entanglement properties are carried out for simple examples.
\end{abstract}

\maketitle 

One of the most fascinating aspects of the two-dimensional systems is the possibility of anyonic statistics, that is both theoretically important and with promising practical applications\cite{Myrheim,wilczek1,wilczek2,mr,kitaev,nayak}. The fractional quantum Hall (FQH) effect, realised by subjecting a two-dimensional electron gas to a strong perpendicular magnetic field, is an ideal platform for anyon fluids\cite{prange}. The possibility of anyons and non-abelions hosted by gapped topological phases was proposed in\cite{laughlin, mr, halperin,stern}, with tentative experimental signals in a number of recent works\cite{banerjee,clarke,manfra}. Even in simple FQH phases, there can be rich dynamics involving the interaction and transmutation between different types of anyons\cite{ajit,yang}.

Theoretically, the statistics and dynamics of anyons can be understood in different ways. Haldane's generalised Pauli exclusion principle extends the notion of bosons and fermions by looking at the reduction of Hilbert space when a particle occupies a state\cite{haldane1}. In this perspective, a ``hard-core" boson and a fermion are equivalent. It is however not apparent if all of the statistical aspects (e.g. the complex phases from adiabatic braiding) are captured within this formalism. Fundamentally, the statistical properties of anyons can be understood as complex interactions between particles. For example, statistical transmutation with flux attachment and various schemes of boson-fermion dualities have been proposed\cite{rr,zee,ragnu,jensen,mulligan,son,fradkin,seidel}. Such dualities can be established if there is an exact mapping of the energy spectrum or partition function from one system to another. It is, however, not easy to understand the ``statistical interaction" beyond the mean-field level in the field theoretical description with flux attachment or singular gauge transformation.

In this Letter, we use anyons in FQHE as the example and propose exact duality not only between bosons and fermions, but also between bosons and anyons. The statistical interaction between anyons can be microscopically derived order by order, shown to be equivalent to the few-body interactions between bosons in the dual description. The ability to bosonize anyons in two dimensions could be understood as a consequence of bulk-edge correspondence\cite{haldane,yan,anushya} and the chiral Luttinger liquid description\cite{chang,wen} of the quantum Hall edge. It is bosonization of the \emph{entire} Hilbert space with conformal symmetry and without non-abelian parafermions\cite{rr,cooper,jackson,santos}, in contrast to the approach with Jordan-Wigner transformation on lattice systems\cite{kapustin}. A direct consequence of this bosonization scheme is a large family of bosonic FQH phases, with explicit model Hamiltonians that in principle can be constructed exactly from the corresponding fermionic FQH phases. These new topological phases are dual descriptions of the familiar FQH phases of the electrons, though they occur at different filling factors and topological shifts.

{\it A bosonic description for fermions--}   The concepts of this work are most conveniently illustrated on the spherical geometry\cite{sphere,geometry} with rotational symmetry. For gapped topological phases, rotational symmetry can also be relaxed, since FQH topological orders do not require any symmetry protection\cite{yang1,yang2}. Without loss of generality we focus on spinless electrons in the lowest Landau level (LLL). With a monopole of total magnetic flux $2\bm S$ at the center of the sphere, the number of single particle states (or orbitals) in the LLL is $N_o=2\bm S+1$. 

It is useful to define the vacuum in this Hilbert space as the highest density state $|\psi_{\mathcal H}\rangle$ of a particular sub-Hilbert space $\mathcal H$ within the LLL. Basis states of interest are thus created from the vacuum by inserting the magnetic fluxes. For example, if $\mathcal H$ is the full Hilbert space of the LLL, then the vacuum is the fully filled LL. All other basis in $\mathcal H$ can be obtained by flux insertion, or the creation of holes, from the vacuum. The vacuum and one such basis are shown as follows:
\begin{eqnarray}\label{basis}
&&11111\cdots 111,\qquad 011111\cdots 111\label{v1}
\end{eqnarray}
where each digit represents an orbital arranged sequentially from the north pole to the south pole on the sphere, with ``1" indicating the orbital is occupied by an electron, and ``0" otherwise. The basis on the right has one hole at the north pole. In both cases, electrons and holes are fermions, with single particle orbitals that are eigenstates of the angular momentum operator $\hat L_z$ and $\hat L^2$. They thus behave like spinors, in the sense that each electron or hole can be represented as a spinor with total angular momentum, or spin $\bm S$, and the quantum state can be labeled as $|\bm s,\bm S\rangle$, where $\bm s=-\bm S, -\bm S+1,\cdots,\bm S-1,\bm S$ is the index of the single particle orbitals.

The fermionic nature of the electrons or holes manifests from the $\hat L^2$ eigenstates for two particles. With more than one particle, we define $\hat L_{\alpha}=\sum_i\hat L_{\alpha,i}, \hat L^2=\hat L_x^2+\hat L_y^2+\hat L_z^2$, where $\alpha=x,y,z$ and $i$ is the index of \emph{electrons}. The total angular momentum $\bm S_{\text{tot}}$ of the two particles can be any integer between $0$ and $2\bm S$. For fermions, however, only $\bm S_{\text{tot}}=2\bm S-k$ with odd integer $k$ is allowed. In contrast for bosons, $k$ can only be even. The counting, or the number of $\hat L^2$ eigenstates for each $\bm S_{\text{tot}}$ with more than two particles\cite{supp}, are also different between fermions and bosons. Such counting is the \emph{signature} of the particle statistics. 

We now show that the fermionic holes in a single LL can be ``bosonized". Starting with a fully filled LL with $N_e$ electrons and as the vacuum for the holes, an insertion of one magnetic flux creates a single hole with total spin $\bm S_{1,h}=N_e/2$. If two magnetic fluxes are inserted, we create two holes each with spin $\bm S'_{1,h}=\left(N_e+1\right)/2$. The total spin of the two holes are thus $\bm S_{2,h}=N_e+1-k'$ with $k'$ odd. The key observation here is that if instead of treating each hole as a fermion with spin $\bm S'_{1,h}$, we can also treat it as a particle with spin $\bm S_{1,h}$, so that the total spin is $\bm S_{2,h}=N_e-k$ with k even: the holes behave like bosons. This applies for the insertion of multiple fluxes with a fixed number of $N_e$: as fermions, each hole has $\bm S'_{1,h}$ that depends on the number of fluxes inserted, but they are also bosons with a fixed $\bm S_{1,h}$ that only depends on $N_e$. In fact, when a fully filled LL with fixed $N_e$ is defined as the vacuum, the inserted fluxes should be understood as ``particles" of spin $\bm S_{1,h}$, since in this way all holes have the same spin, independent of the number of fluxes inserted.

This seemingly trivial reinterpretation has important consequences. If we insert $N_h$ fluxes to the fully filled LL, with the number of orbitals $N_o=N_e+N_h$, the Hilbert space is spanned by fermionic product states $|\bm s'_1,\bm S'_{1,h};\bm s'_2,\bm S'_{1,h};\cdots;\bm s'_{N_h},\bm S'_{1,h}\rangle$ with $\bm S'_{1,h}=\left(N_o-1\right)/2$ and $\bm s'_i$ running from $-\bm S'_{1,h}$ to $\bm S'_{1,h}$. The same space is also spanned by the bosonic product states $|\bm s_1,\bm S_{1,h};\bm s_2,\bm S_{1,h};\cdots;\bm s_{N_h},\bm S_{1,h}\rangle$ with $\bm S_{1,h}=N_e/2$ and $\bm s_i$ running from $-\bm S_{1,h}$ to $\bm S_{1,h}$, where each bosonic product state is a linear combination of the fermionic counterpart (so an entangled many-body state of fermionic holes, see\cite{supp}). Thus all physics in a single LL can either be understood from quantum states of fermions (with $N_e$ electrons and $N_o$ orbitals), or quantum states of bosons (with $N_o-N_e$ bosons, and $N_e+1$ orbitals). The two Hilbert spaces have the same dimension.

{\it Bosonization of anyons--} This bosonic description of the Hilbert space can be easily generalised to anyons. An insertion of the magnetic flux creates a fermion in the full Hilbert space, but it will create an anyon in the truncated Hilbert space $\tilde{\mathcal H}\in\mathcal H$. The Hilbert space truncation can either be implemented via model Hamiltonians\cite{simon}, or more generally using LEC\cite{yang3}. Each truncated Hilbert space corresponds to a topological FQHE phase, which we can index with the filling factor $\nu$ and the topological shift $S$\cite{wenzee,readshift}. Note we have the relationship $N_o=\nu^{-1}N_e-S$ for the ground state, which serves as the vacuum. The previously discussed full Hilbert space and its fully filled LL as the vacuum can thus be denoted as $\mathcal H_{[\nu,S]}=\mathcal H_{[1,0]}$ and $|\psi\rangle_{[\nu,S]}=|\psi\rangle_{[1,0]}$.

Let us illustrate the duality between anyons and bosons with the simple example of the Laughlin phase at filling factor $\nu=1/3$. The universal topological properties of this phase are defined by the null space of the Haldane pseudopotential interaction $\hat V^{\text{2bdy}}_1$, denoted as $\mathcal H_{[\frac{1}{3},-2]}$, spanned by the exact zero energy states\cite{prange}, or Laughlin ground states and quasihole states. Thus the vacuum, or the highest density state for a given $N_e$, is the Laughlin ground state denoted as $|\psi\rangle_{[\frac{1}{3},-2]}$. Insertion of the magnetic fluxes creates anyons of charge $e/3$, instead of the fermionic holes in the full Hilbert space.

It is natural to organise the Laughlin quasiholes into eigenstates of $\hat L^2$ and $\hat L_z$, and the counting of these eigenstates clearly indicates the quasihole subspace is spanned by the bosonic degrees of freedom. For a single Laughlin quasihole, it has total angular momentum $\bm S_{1,\text{qh}}=N_e/2$, same as the fermionic holes. They are many-body wavefunctions in the electron basis denoted as $|\bm s,\bm S_{1,\text{qh}}\rangle$ with $\bm s=-N_e/2,-N_e/2+1,\cdots, N_e/2$. We also know they are Jack polynomials, and using $N_e=4$ as an example, the root configurations of the five single-quasihole states are\cite{bernevig}:
\begin{eqnarray}
&&01001001001, 10001001001, 10010001001\nonumber\\
&&\qquad\quad 10010010001, 10010010010
\end{eqnarray}
Inserting a second magnetic flux to those root configurations creates the two-quasihole states. 
By diagonalising the quasihole subspace with $\hat L^2$ we see it is more natural to treat these quasiholes as bosons, each with $\bm S_{1,\text{qh}}=N_e/2$, independent of the total number of orbitals. Same as the fermionic holes, for the Hilbert space with $N_e$ electrons and $N_{\text{qh}}$ Laughlin quasiholes, we can construct an orthonormal basis denoted as $|\bm s_1,\bm S_{1,\text{qh}};\bm s_2,\bm S_{1,\text{qh}};\cdots;\bm s_{N_{\text{qh}}},\bm S_{1,\text{qh}}\rangle$ with $\bm s_i$ running from $-\bm S_{1,\text{qh}}$ to $\bm S_{1,\text{qh}}$. Each state is a strongly entangled state in the electron basis, but can be interpreted as a ``product state" in the bosonic quasihole basis\cite{supp}. Again the two descriptions are equivalent in $\mathcal H_{[\frac{1}{3},-2]}$. One should note that for two localised Laughlin quasiholes far apart from each other, braiding one quasihole around the other leads to anyonic Berry phase. What we have shown here is that for the many-quasihole states, there are always proper linear combinations of them that give states containing particles behaving like \emph{bosons}, just like the case with the fermionic holes in $\mathcal H_{[1,0]}$.
\begin{figure}
\begin{center}
\includegraphics[width=\linewidth]{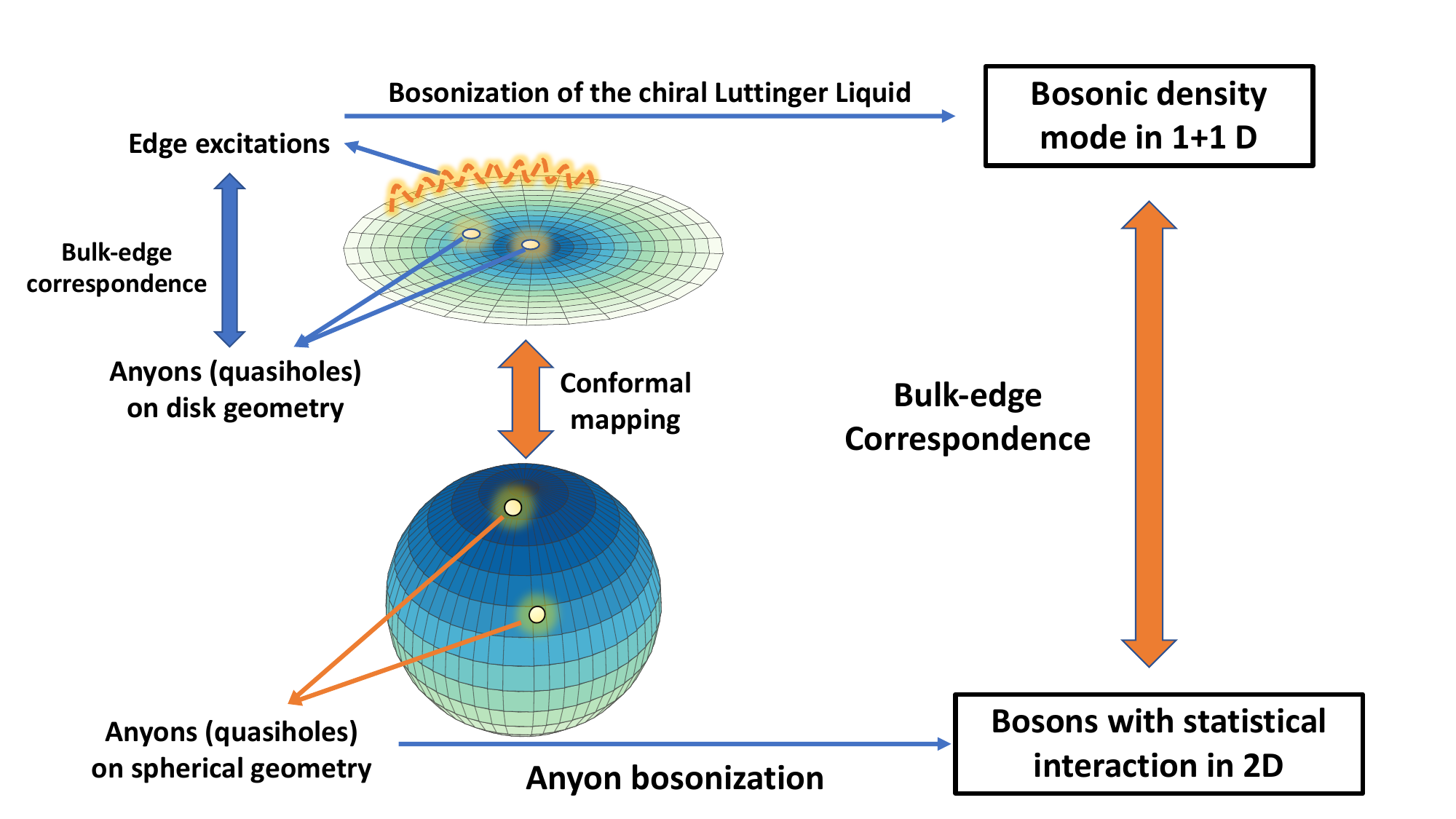}
\caption{A schematic illustration of the bosonization of anyons in 2D.}
\label{fig1}
\end{center}
\end{figure}

The bosonization of anyons can be applied to the null spaces of model Hamiltonians of other abelian FQH phases, as well as Hilbert spaces defined by LEC, as long as the counting of quasiholes is abelian. It can also be applied to the Hilbert space spanned by the ground state and quasihole states of the abelian composite fermion (CF) states, as long as $\mathcal H_{[\nu,S]}$ and $|\psi\rangle_{[\nu,S]}$ can be properly constructed from the CF theory by mapping the FQH states to the IQH states of CFs (with the built-in assumption of bulk-edge correspondence)\cite{jain,ajit2,toke,ywz}, even in the absence of an exact model Hamiltonian. Fundamentally, this possibility of bosonization in two dimensions is due to the conformal mapping and bulk-edge correspondence of the FQH fluids. All quasihole states on the spherical geometry can be conformally mapped to the disk geometry, where the insertion of the magnetic fluxes in the bulk is equivalent to edge excitations at the boundary\cite{read}. Bosonic representation of the anyons in the bulk (even for geometries without boundaries) is thus the dual description of the density modes of the edge of the quantum Hall fluids(see Fig.(\ref{fig1})). However, if $\mathcal H_{[\nu,S]}$ is non-Abelian, one cannot fully bosonize the quasiholes within $\mathcal H_{[\nu,S]}$ due to the presence of parafermions\cite{rr,cooper,jackson,santos}.

{\it Bosonic topological phases--} For non-interacting electrons in a single LL with no disorder, the Hamiltonian is just an identity. In the dual description, even though the statistical properties of the particles have changed, the bosons are also non-interacting. Similarly, if we bosonize within $\mathcal H_{[\frac{1}{3},-2]}$ when electrons interact with $\hat V^{\text{2bdy}}_1$ (so that Laughlin quasiholes are non-interacting), the resulting bosons will be non-interacting as well. Thus in a system with no dynamics at all, statistical interaction is also absent. If we introduce interaction between electrons or anyons, non-trivial interaction between bosons in the dual picture will develop. The latter can be determined by imposing exact mapping of the energy spectra in the two pictures, in addition to the mapping of the many-body states we have established. The microscopic interaction between bosons then also captures the statistical interaction between fermions or anyons in the respective (truncated) Hilbert space. 

We illustrate this with the full Hilbert space $\mathcal H_{[1,0]}$, and introduce $\hat V^{\text{2bdy}}_1$ between electrons. For $N_e$ electrons and $N_h$ holes, we can label the eigenstates of $\hat V^{\text{2bdy}}_1$ with $|\bm S,\alpha, N_h\rangle_h$ and the respective energy $E_{\bm S,\alpha,N_h}$. Only the highest weight states are needed so that $\hat L_z|\bm S,\alpha,N_h\rangle_h=\bm S|\bm S,\alpha,N_h\rangle_h$ and $\hat L^2|\bm S,\alpha,N_h\rangle_h=\bm S\left(\bm S+1\right)|\bm S,\alpha,N_h\rangle_h$, with $\alpha$ the index labelling the degeneracy of the highest weight states in each total angular momentum sector. We now know that for a bosonic Hilbert space with $N_h$ bosons and $N_e+1$ orbitals, each $|\bm S,\alpha,N_h\rangle_h$ have a one-to-one mapping to a bosonic state $|\bm S,\alpha,N_h\rangle_b$ with the same quantum numbers. The effective Hamiltonian between the bosons is given as follows:
\begin{eqnarray}\label{hb1}
\hat H_b=\sum_{n=2}^{\infty}\sum_{k,\alpha_k}\lambda_{\bm S_k,\alpha_k,n}\hat V_{k,\alpha_k}^{\text{n-bdy}}
\end{eqnarray}
where $\bm S_k=\bm S-k$, $\hat V_{k,\alpha_k}^{\text{n-bdy}}$ are the n-body pseudopotentials\cite{simon} with total relative angular momentum $k$, and $\alpha_k$ labels the degeneracy of the pseudopotentials. We thus require
\begin{eqnarray}\label{iteration}
{}_b\langle\bm S_k,\alpha_k,N_h|\hat H_b|\bm S_k,\alpha_k,N_h\rangle_b=E_{\bm S_k,\alpha_k,N_h}
\end{eqnarray}
Using different values of $N_h$, the coefficients of $\lambda_{\bm S_k,\alpha_k,n}$ can be computed iteratively\cite{supp}. For example, it is easy to check that $\lambda_{N_e,1,2}=1$, and $\lambda_{N_e-2p,1,2}=0$ for integer $p>0$. In the thermodynamic limit, we can thus show analytically the following\cite{supp}:
\begin{eqnarray}
&&\lambda_{\bm S_n,1,n}=E_{\bm S_n,1,n}-\sum_{k'=2}^{n-1}\lambda_{\bm S_{k'},1,k'}\frac{n!}{\left(n-k'\right)!k'!}\qquad\\
&&E_{\bm S_n,1,n}=\sum_{m',n'=0}^{n-1}\frac{\left(m'+n'-1\right)!}{2^{m'+n'-1}m'!n'!}\left(m'-n'\right)^2
\end{eqnarray}
where $\bm S_n=\frac{nN_e}{2}$. Here $\lambda_{\bm S_n,1,n}$ is the coefficient of the leading n-body bosonic pseudopotential with \emph{zero} total relative angular momentum. In general, $\lambda_{\bm S_n-2p,1,n}$ \emph{decreases rapidly} with increasing $p$, but rather slowly with increasing $n$ (see\cite{supp}). Thus while the interaction between fermionic holes is just a short-range two-body interaction, the interaction between the bosons in the dual picture is longer-ranged, with few-body interactions involving clusters of bosons. The complexity of the bosonic interaction reflects the statistical interaction between the underlying fermions, which we can now quantify order by order.

\begin{figure}
\begin{center}
\includegraphics[width=\linewidth]{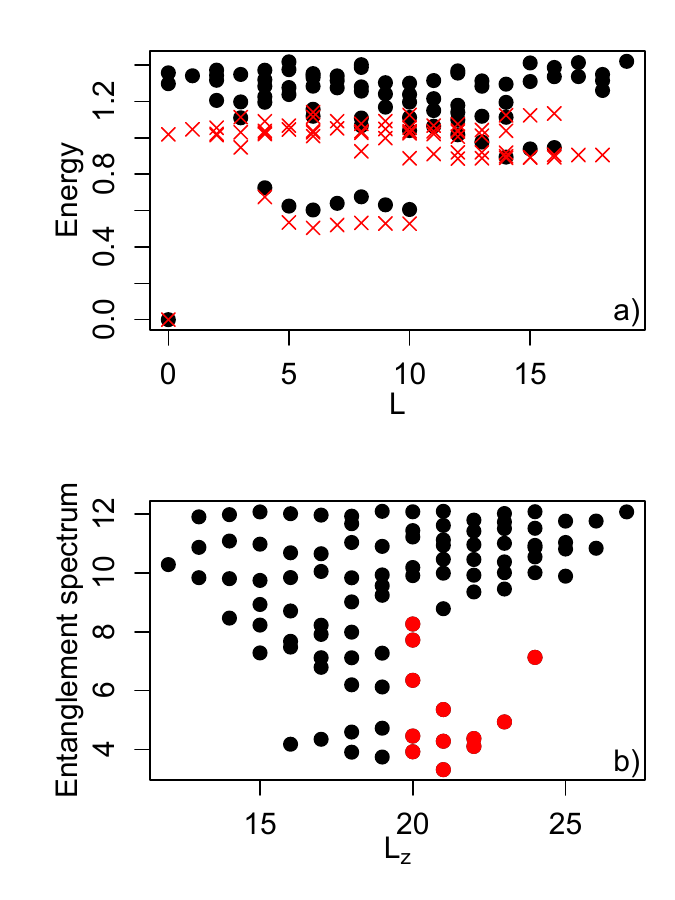}
\caption{a). The black dots give the energy spectrum with $18$ bosons, $11$ orbitals, and a truncated Hamiltonian dual to $\hat V_1^{\text{2bdy}}$ as detailed in\cite{supp}, and the fermionic spectrum of $\hat V_1^{\text{2bdy}}$ is shown here with red crosses as comparison. b). The ground state entanglement spectrum, with low-lying states (shown in red) having the correct Virasoro counting.}
\label{fig2}
\end{center}
\end{figure}

If we choose $N_h=2\left(N_e-1\right)$, we are at the filling factor $\nu=1/3$ with $S=-2$, and $\hat V^{\text{2bdy}}_1$ gives the familiar Laughlin phase. In the bosonic picture, the corresponding filling factor is $\nu=2$ with $S=2$. We will thus obtain a previously unreported gapped topological phase with the effective interaction $\hat H_b$, that has the exact same spectrum as the fermionic Laughlin phase. It can potentially be realised in experiments, because while $\hat H_b$ seems very artificial and hard to engineer experimentally, for gapped phases one can have realistic interaction adiabatically connected to the model interactions (an example is the MR state in the second LL\cite{sarma}). In Fig.(\ref{fig2}) we show that with a much simpler truncated $\hat H_b$, the bosonic spectrum highly resembles the fermionic topological phase, and low-lying states in the ground state entanglement spectrum captures the Laughlin edge modes as expected.

We can also define more than one vacuum for the same topological phase. For example, if we are interested in the Laughlin phase at $\nu=1/5$, we can treat the states as quantum fluids of holes created from $|\psi\rangle_{[1,0]}$. This is always possible for any quantum fluids in the LLL. In the Hilbert space of $N_o=5N_e-4$, bosonization of the fermionic holes maps it to the bosonic system with $N_e+1$ orbitals and $4\left(N_e-1\right)$ bosons, corresponding to a filling factor of $\nu=4$ and $S=2$, with the effective interaction Hamiltonian $\hat H_b$ defined in Eq.(\ref{hb1}) and computed from Eq.(\ref{iteration}). Note that $E_{\bm S_k,\alpha_k,N_h}$ in Eq.(\ref{iteration}) needs to be computed from the bare interaction from the electrons, i.e. the Haldane pseudopotentials $\hat V_1^{\text{2bdy}}+\hat V_3^{\text{2bdy}}$ for the case of the model Hamiltonian at $\nu=1/5$.

Alternatively, we can treat the states as quantum fluids of the Laughlin $\nu=1/3$ quasiholes created from $|\psi\rangle_{[\frac{1}{3},-2]}$. 
Bosonization of the same topological phase as quantum fluids of Laughlin $\nu=1/3$ quasiholes leads again to the bosonic system with $N_e+1$ orbitals and $2\left(N_e-1\right)$ bosons, corresponding to a filling factor of $\nu=2$ and $S=2$. This is the same as the bosonic description of the Laughlin $\nu=1/3$ phase in $\mathcal H_{[1,0]}$. These could be two competing bosonic phases at the same filling factor and topological shift, 
 each with a well-defined model Hamiltonian. It would be interesting to see if these two Hamiltonians are topologically distinct, given that they are very different with the pseudopotential expansion\cite{supp}, and are the dual description of the fermionic Laughlin phase at different filling factors. The topological entanglement entropy\cite{preskill,levin} could be the topological index that distinguishes between the two bosonic phases. 
 
{\it Summary and outlook--}
We have established that \emph{all} fractional quantum Hall fluids (including non-abelian ones) can be understood as quantum fluids of bosons, when anyons (including fermionic holes)  are bosonized either in the single LL Hilbert space $\mathcal H_{[1,0]}$, or other abelian sub-Hilbert spaces. 
From the bare interaction between electrons, this bosonization scheme allows us to explicitly calculate the statistical interaction between anyons, and construct microscopic Hamiltonians for the dynamics of the anyon fluids. The duality hints that each quantum Hall fluid can be understood equivalently as particles of different statistical properties. This may explain various different effective schemes involving composite fermions or bosons in the literature as viewing the same physics from different perspectives\cite{hlr,sondirac,dx,shankar,ajit3}.

The bosonization scheme allows us to identify topological FQH phases of interacting bosons with the topological orders inherent from their fermionic/anyonic counterparts. They are different from the two-component or lattice bosonic FQHE proposed in the literature\cite{senthil,vish,yinghai}, that require symmetry protection. A family of fermionic parton states was recently proposed in\cite{ajit}, and it is possible that their bosonic versions could be related to the topological phases proposed here\cite{private}. We numerically analysed the gapped energy spectrum and the ground state entanglement spectrum of a bosonic phase at $\nu=2$ and $S=2$, confirming the validity of the bosonization scheme. Note that one fractional quantum Hall phase for electrons can be interpreted either as a quantum fluid of fermionic holes, or different types of anyons. Each interpretation leads to a dual description of bosons with different microscopic interactions, at different filling factors. It will be interesting to see if we can experimentally realise such bosonic FQH phases in, for example, the cold atom systems; or to see how these bosonic excitations with fractional charges can be nucleated in the bulk of the fermionic FQH systems.

Fundamentally, the ability to bosonize anyons in two-dimensions is related to the bulk-edge correspondence and the feasibility of bosonizing the chiral Luttinger liquid of the edge excitations. For FQH phases with ground states and quasiholes living in non-abelian subspaces (e.g. $\mathcal H_{[\frac{1}{2},-2]}$ with vacuum $|\psi\rangle_{[\frac{1}{2},-2]}$, which is the null space of the Moore-Read model three-body Hamiltonian), we cannot fully bosonize the quantum Hall fluid, because of the presence of parafermions\cite{footnote}. Nevertheless, we can still understand the ground state and quasiholes of such FQH phases as quantum fluids of bosons and parafermions, with explicit statistical interactions between them from a similar scheme. Further studies of such systems could enhance our understandings of the statistical nature of anyons and non-abelions, and help to construct effective field theory descriptions of such exotic particles in a systematic manner.

\begin{acknowledgments}
{\sl Acknowledgements.} I thank A.C. Balram for useful discussions, H.Q. Trung for helping me improve the figures, and Y.Z. Wang for pointing me to the partition theory for the state counting on the sphere. This work is supported by the NTU grant for Nanyang Assistant Professorship and the National Research Foundation, Singapore under the NRF fellowship award (NRF-NRFF12-2020-005).
\end{acknowledgments}



\clearpage

\renewcommand{\thefigure}{S\arabic{figure}}
\renewcommand{\theequation}{S\arabic{equation}}
\renewcommand{\thepage}{S\arabic{page}}
\setcounter{figure}{0}
\setcounter{page}{1}

\onecolumngrid
\begin{center}
\textbf{\large Supplementary Online Materials for ``Statistical interactions and boson-anyon duality in fractional quantum Hall fluids"}
\end{center}
\setcounter{equation}{0}
\setcounter{figure}{0}
\setcounter{table}{0}
\setcounter{page}{1}
\makeatletter
\renewcommand{\theequation}{S\arabic{equation}}
\renewcommand{\thefigure}{S\arabic{figure}}
\renewcommand{\bibnumfmt}[1]{[S#1]}
\renewcommand{\citenumfont}[1]{S#1}
In this supplementary material, we give more technical details on the duality between the anyon and boson description of the same Hilbert space of the quantum fluids in a single Landau level (LL). We also show examples of the explicit construction of the interaction Hamiltonians for bosons, that captures the statistical interaction between anyons.

\section{S1. Bosonic product states from anyonic many-body wavefunctions}

Let us look at a specific example with a model Hamiltonian as follows:
\begin{eqnarray}
\hat H_a=\hat V^{\text{2bdy}}_1+\hat V^{\text{2bdy}}_3
\end{eqnarray}
where $\hat V^{\text{2bdy}}_1,\hat V^{\text{2bdy}}_3$ are the two leading two-body Haldane pseudopotential interaction, so $\hat H_a$ is the model Hamiltonian of the Laughlin $\nu=1/5$ phase. We now have a well-defined Hilbert space $\mathcal H_a$ as the null space of $\hat H_a$. This Hilbert space contains the Laughlin $\nu=1/5$ ground state, as well as its quasiholes. These quasiholes are anyons of charge $e/5$. For $|\psi\rangle\in\mathcal H_a$, we know that $N_o\ge 5N_e-4$ on the sphere, where $N_e$ is the number of electrons and $N_o$ is the number of orbitals.

We would like to bosonize $\mathcal H_a$ within another Hilbert space $\mathcal H_b\supseteq \mathcal H_a$. Note that this can be done even if $\mathcal H_a$ is non-abelian (i.e. it works if $\mathcal H_a$ is the null space of the Moore-Read model Hamiltonian, etc.), as long as $\mathcal H_b$ is abelian. As discussed in the main text, there can be different choices of $\mathcal H_b$, depending on which type of anyons we would like to bosonize. We will go through several examples here.

Since in this case $\mathcal H_a$ is also abelian, we can consider the case where $\mathcal H_b=\mathcal H_a=\mathcal H_{[\frac{1}{5},-4]}$, using the notations introduced in the main text, so that the vacuum is the Laughlin $\nu=1/5$ ground state, which we can denote as $|\psi\rangle_{[\frac{1}{5},-4]}$. The Hilbert space of $\mathcal H_a$ is thus generated by inserting magnetic fluxes into $|\psi\rangle_{[\frac{1}{5},-4]}$. Let the number of inserted fluxes be $N_{\text{qh}}$, so for fixed $N_e$ and $N_{\text{qh}}$ we have $N_o=5N_e-4+N_{\text{qh}}$. For this subspace of $\mathcal H_a$, which we denote as $\mathcal H_{a,N_e,N_o}$, it is spanned by fermionic Jack polynomials that are zero energy states of $\hat H_a$, though we do not need to use this aspect explicitly. 

As argued in the main text, we can treat each one of the $N_{\text{qh}}$ quasiholes as a spinor with total spin $\bm S_1=N_e/2$. By diagonalising the total angular momentum operator $\hat L^2$ within $\mathcal H_{a,N_e,N_o}$, we can label the highest weight eigenstates as follows:
\begin{eqnarray}
\hat L^2|\bm S_k,\bm S_k,\alpha_k\rangle_a=\bm S_k\left(\bm S_k+1\right)|\bm S_k,\bm S_k,\alpha_k\rangle_a, \qquad\hat L_z|\bm S_k,\bm S_k,\alpha_k\rangle_a=\bm S_k|\bm S_k,\bm S_k,\alpha_k\rangle_a
\end{eqnarray}
Here $\bm S_k=N_{\text{qh}}N_e/2-k$, the first quantum number is related to $\hat L_z$, the second quantum number is related to $\hat L^2$, and $\alpha_k$ labels the states with total relative angular momentum $k$. Let $\bar\alpha_{k,N_{\text{qh}}}$ be the degeneracy of states with total relative angular momentum $k$, the bosonic nature of the quasiholes is reflected by $\bar\alpha_{k,N_{\text{qh}}}$ as shown in Table \ref{t1}.  In contrast for fermions, we need to shift $k$ to $k+N_{\text{qh}}\left(N_{\text{qh}}-1\right)/2$. Each of the eigenstate $|\bm S_k,\bm S_k,\alpha_k\rangle_a$ is a many-body wavefunction of electrons.
\begin{table}[h!]
\centering
\begin{tabular}{|c|c|c|c|c|c|c|} 
 \hline
  & $N_{\text{qh}}=2$ & $N_{\text{qh}}=3$& $N_{\text{qh}}=4$ & $N_{\text{qh}}=5$ & $N_{\text{qh}}=6$&$N_{\text{qh}}=7$\\ 
 \hline
$k=0$  & $1$ & $1$ & 1 & 1&1 & 1\\ 
 \hline
$k=1$&   $0$ & $0$ & 0 & 0&0 & 0 \\ 
 \hline
$k=2$ &  $1$ & $1$ & 1 & 1&1 & 1 \\ 
 \hline
 $k=3$ &  $0$ & $1$ & 1 & 1&1 & 1 \\ 
 \hline
 $k=4$ &  $1$ & $1$ & 2 & 2&2 & 2 \\ 
 \hline
 $k=5$ &  $0$ & $1$ & 1 & 2&2 & 2 \\ 
 \hline
  $k=6$ &  $1$ & $2$ & 3 & 3&4 & 4 \\ 
 \hline
  $k=7$ &  $0$ & $1$ & 2 & 3&3 & 4 \\ 
 \hline
  $k=8$ &  $1$ & $2$ & 4 & 5&6 & 6 \\ 
 \hline
  $k=9$ &  $0$ & $2$ & 3 & 5&6 & 7 \\ 
 \hline
  $k=10$ &  $1$ & $2$ & 5 & 7&9 & 10 \\ 
 \hline
  $k=11$ &  $0$ & $2$ & 4 & 7&9 & 11 \\ 
 \hline
  $k=12$ &  $1$ & $3$ & 7 & 10&14 & 16 \\ 
 \hline
\end{tabular}
\caption{Some values of the degeneracy of the total angular momentum sector $\bar\alpha_{k,N_{\text{qh}}}$, for different values of $N_{\text{qh}}$ and $k$.}
\label{t1}
\end{table}

We now show how to construct bosonic product states as a linear combination of $|\bm s_k,\bm S_k,\alpha_k\rangle$, where $|\bm s_k,\bm S_k,\alpha_k\rangle\sim \left(\hat L_{-}\right)^{\bm S_k-\bm s_k}|\bm S_k,\bm S_k,\alpha_k\rangle$. The dual Hilbert space to the anyonic $\mathcal H_{a,N_e,N_o}$ (with $N_o=5N_e-4+N_{\text{qh}}$)  is the bosonic Hilbert space $\mathcal H_{b,N_{\text{qh}},N_e+1}$, containing $N_{\text{qh}}$ bosons and $N_e+1$ orbitals. In this picture, each boson is also a spinor with total spin $\bm S_1=N_e/2$, and we can diagonalise $\hat L^2$ within this bosonic space to get the following eigenstates:
\begin{eqnarray}
\hat L^2|\bm S_k,\bm S_k,\beta_k\rangle_b=\bm S_k\left(\bm S_k+1\right)|\bm S_k,\bm S_k,\beta_k\rangle_b, \qquad\hat L_z|\bm S_k,\bm S_k,\beta\rangle_b=\bm S_k|\bm S_k,\bm S_k,\beta_k\rangle_b
\end{eqnarray}
again with $\bm S_k=N_{\text{qh}}N_e/2-k$, and $\beta_k$ labels the states with the same total relative angular momentum. Note that the degeneracy of the bosonic states $\bar\beta_{k,N_{\text{qh}}}=\bar\alpha_{k,N_{\text{qh}}}$ and we have a one-to-one mapping between $|\bm s_k,\bm S_k,\beta_k\rangle_b$ and $|\bm s_k,\bm S_k,\alpha_k\rangle_a$. The monomials in $\mathcal H_{b,N_{\text{qh}},N_e+1}$ can be denoted as $|m_0,m_1,\cdots m_{N_e}\rangle_b$, where $m_i$ is the number of bosons in the $i^{\text{th}}$ orbital (corresponding to the $\hat L_z$ quantum number $\bm S_1-i$), with $\sum_im_i=N_{\text{qh}}$. One can thus easily express these product states as follows:
\begin{eqnarray}
|m_0,m_1,\cdots m_{N_e}\rangle_b=\sum_{k=0,\beta_k=1}^{k=N_{\text{qh}}N_e/2,\beta_k=\bar\beta_{k,N_{\text{qh}}}}\lambda_{k,\beta_k}|\bm s_k,\bm S_k,\beta_k\rangle_b\label{b}
\end{eqnarray}
with $\bm m_k=\sum_{i=0}^{N_e}m_i\left(\frac{N_e}{2}-i\right)$, and $\lambda_{k,\beta_k}$ are related to the Clebsch-Gordon coefficients and can be explicitly found. Using the same coefficients of $\lambda_{k,\beta_k}$, we can thus define the following basis as linear combinations of the many-body wavefunctions in the electron basis:
\begin{eqnarray}
|m_0,m_1,\cdots m_{N_e}\rangle_a&=&\sum_{k=0,\alpha_k=1}^{k=N_{\text{qh}}N_e/2,\alpha_k=\bar\beta_{k,N_{\text{qh}}}}\lambda_{k,\alpha_k}|\bm s_k,\bm S_k,\alpha_k\rangle_a\label{b1}
\end{eqnarray}
and one can check the following two relations hold:
\begin{eqnarray}\label{raise}
\hat L_+|m_0,m_1,\cdots m_{N_e}\rangle_a&=&\sum_i\sqrt{m_i\left(m_{i+1}+1\right)}\sqrt{\left(N_e-i\right)\left(i+1\right)}|m_0,\cdots, m_i-1,m_{i+1}+1,\cdots,m_{N_e}\rangle_a\label{b2}\\
\hat L_-|m_0,m_1,\cdots m_{N_e}\rangle_a&=&\sum_i\sqrt{m_i\left(m_{i-1}+1\right)}\sqrt{\left(N_e-i+1\right)i}|m_0,\cdots, m_{i-1}+1,m_{i}-1,\cdots,m_{N_e}\rangle_a\label{b3}
\end{eqnarray}
From Eq,(\ref{b2}) and Eq.(\ref{b3}), we see that $|m_0,m_1,\cdots m_{N_e}\rangle_a$ indeed gives an orthonormal basis behaving like bosons, and can be treated as a monomial of bosons. Note given that $\hat L_+,\hat L_-$ act on electrons, they are effectively $\hat L_-,\hat L_+$ of the bosons, since the bosons are fundamentally from the bosonization of the magnetic fluxes.

The bosonization scheme with a different $\mathcal H_b$ follows the same procedures. If we take $\mathcal H_b$ to be the null space of $\hat V^{\text{2bdy}}_1$, then we have the Laughlin $\nu=1/3$ ground state, $|\psi\rangle_{[\frac{1}{3},-2]}$ as the vacuum. In this subspace, not only can $\mathcal H_a$ be bosonized, any zero energy state of $\hat V^{\text{2bdy}}_1$ with $3N_e-2<N_o$ can be bosonized. In particular, the relationships in Eq.(\ref{b}) - Eq.(\ref{b3}) still follow, though $N_{\text{qh}}=N_o-\left(3N_e-2\right)$ (while in $\mathcal H_{[\frac{1}{5},-4]}$ we have $N_{\text{qh}}=N_o-\left(5N_e-4\right)$, since we have a different vacuum). Moreover, $|\bm s_k,\bm S_k,\alpha_k\rangle_a$ will be different, since they are eigenstates of $\hat L^2$ in $\mathcal H_{[\frac{1}{3},-2]}$ instead of $\mathcal H_{[\frac{1}{5},-4]}$. We can also use a different $\mathcal H_b=\mathcal H_{[1,0]}$, which is the full Hilbert space with the completely filled LL as the vacuum. In all three cases, the bosonic monomials $|m_0,m_1,\cdots m_{N_e}\rangle_a$ are different, so we are dealing with three different types of bosons as the dual description of the anyons.

One should also note that the mapping from $|\bm s_k,\bm S_k,\alpha_k\rangle_a$ to $|\bm s_k,\bm S_k,\beta_k\rangle_b$ is not unqiue. Any unitary transformation of $|\bm s_k,\bm S_k,\beta_k\rangle_b$ in the subspace of fixed $\bm s_k,\bm S_k$ (so the subspace with dimension $\bar\beta_{k,N_{\text{qh}}}$) gives a mapping leading to a complete basis of $|m_0,m_1,\cdots m_{N_e}\rangle_a$ of the many-body wavefunctions that satisfies Eq.(\ref{b2}) and Eq.(\ref{b3}). Any specific mapping can thus be treated as a gauge choice for the statistical transmutation, and a detailed study of this gauge degrees of freedom will be carried out in future works.

Given that the bosonization scheme here can be rigorously defined as a unitary transformation between the fermionic/anyonic state and the bosonic state, the duality also allows us to explicitly write down the relationship between the bosonic and fermionic operators involved in the second quantized language. One should note the bosonic and the fermionic operators have different vacua. As an example, let us look at the Laughlin quasihole states at $\nu=1/3$ with $N_e$ electrons and $N_{\text{qh}}$ quasiholes. This subspace is spanned by bosonic states with $N_{\text{qh}}$ bosons in $N_e+1$ orbitals, and we can thus define the bosonic creation and annihilation operators as $\hat b_i,\hat b_j^\dagger$ with $[\hat b_i,\hat b_j^\dagger]=\delta_{ij}$. The complete basis is given as follows:
\begin{eqnarray}\label{relation}
|n_1,n_2\cdots,n_{N_{\text{qh}}}\rangle=\prod_{1\le i\le N_{\text{qh}}}\hat b_{n_i}^\dagger|\text{vac}\rangle_b=\sum_\lambda c_\lambda|e_\lambda\rangle=\sum_\lambda c_\lambda\prod_{1\le j\le N_e}\hat c^\dagger_{\lambda_j}|\text{vac}\rangle_e
\end{eqnarray}
Here $|n_1,n_2\cdots,n_{N_{\text{qh}}}\rangle$ is just another representation of the many-body state of $|m_0,m_1,\cdots m_{N_e}\rangle$ in Eq.(\ref{raise}), where $n_i$ gives the orbital index of the $i^{\text{th}}$ boson; $|\text{vac}\rangle_b$ is the Laughlin ground state with $N_e$ electrons, serving as the vacuum of the flux insertion. Thus $|n_1,n_2\cdots,n_{N_{\text{qh}}}\rangle$ as a many-body state can be expanded in terms of the electron monomials $|e_\lambda\rangle$, where the coefficients of expansion $c_\lambda$ are unambiguously determined from the bosonization scheme we developed here. For each monomial we have $|e_\lambda\rangle=\prod_{1\le j\le N_e}\hat c^\dagger_{\lambda_j}|\text{vac}\rangle_e$, with $\hat c^\dagger_{\lambda_j}$ creating an electron in the $\lambda_j^{\text{th}}$ orbital, from the electron vacuum $|\text{vac}\rangle_e$, which is an empty LL with the number of orbitals $N_o=3N_e-2+N_{\text{qh}}$ . Eq.(\ref{relation}) determines the relationships between the bosonic $\hat b_i,\hat b_i^\dagger$ and the fermionic $\hat c_i.\hat c_i^\dagger$, explicitly showing that the bosonic operator creates collective excitations involving many electrons from the vacuum of the Laughlin ground state.

\section{S2. Statistical interactions of anyons}

We now show explicitly how the statistical interactions between anyons can be derived as few-body interactions between bosons. Let $\hat H_e$ be the bare interaction between electrons, that is rotationally invariant. This allows us to write $|\bm S_k,\alpha_k,N_{\text{qh}}\rangle_a$ as simultaneous eigenstates of $\hat H_e$, with the eigenenergies $E_{\bm S_k,\alpha_k,N_{\text{qh}}}$. In this section, we only need to deal with the highest weight states, so the $\hat L_z$ quantum number is omitted in the notation. Here the dependence on $N_e$ is also made implicit, and $N_{\text{qh}}$ is the number of quasiholes, or flux insertions to the vacuum. In particular, for $|\bm S_k,\alpha_k,2\rangle_a$ contains two quasiholes, the allowed values of $k$ are $k=0,2,4,\cdots$ and $\bar\alpha_{k,N_{\text{qh}}}=1$. If these two quasiholes are actually holes from the insertion of the magnetic fluxes into a fully filled LL, then the corresponding $E_{\bm S_k,1,2}$ gives the $\left(k+1\right)^{\text{th}}$ two-body pseudopotential components of $\hat H_e$. 

Let the corresponding bosonic effective interaction have the following most general form:
\begin{eqnarray}\label{hb}
\hat H_b=\sum_{n=2}^{\infty}\sum_{\bm S_k,\alpha_k}\lambda_{\bm S_k,\alpha_k,n}\hat V_{k,\alpha_k}^{\text{n-bdy}}
\end{eqnarray}
As explained in the main text, $\hat V_{k,\alpha_k}^{\text{n-bdy}}$ is the $n-$body pseudopotential (PP) with total relative angular momentum $k$, and $\alpha_k$ is the index for the degeneracy of the PP. Clearly $\lambda_{\bm S_k,\alpha_k,n}$ is only non-vanishing for $\bm S_k$ corresponding to bosons. Since we have the exact mapping of the many-body electron wavefunctions $|\bm S_k,\alpha_k,N_{\text{qh}}\rangle_a$ to the bosonic many-body wavefunctions $|\bm S_k,\alpha_k,n\rangle_b$ (which are also eigenstates of $\hat L_2$ in the bosonic Hilbert space), where $n$ is the number of bosons, then we can compute each of $\lambda_{\bm S_k,\alpha_k,n}$. For $n=2$, we immediately have $\lambda_{\bm S_k,1,2}=E_{\bm S_k,1,2}$ for $k=0,2,4,\cdots$.

Now let us look at $n=3$. For the bosonic $\hat L^2$ eigenstates $|\bm S_k,\alpha_k,3\rangle_b$, we have the allowed values of $k=0,2,3,\cdots$ and the corresponding $\bar\alpha_{k,N_{\text{qh}}}$ as listed in Table~\ref{t1}. All values of $\lambda_{\bm S_k,\alpha_k,3}$ can be computed from the following relationship:
\begin{eqnarray}
{}_b\langle\bm S_k,\alpha_k,3|\hat H_b|\bm S_k,\alpha_k,3\rangle_b&=&\sum_{k=0,2,\cdots}\lambda_{\bm S_k,1,2}\cdot{}_b\langle\bm S_k,\alpha_k,3|\hat V_{k,\alpha_k}^{\text{2-bdy}}|\bm S_k,\alpha_k,3\rangle_b+\lambda_{\bm S_k,\alpha_k,3}\nonumber\\
&=&{}_a\langle\bm S_k,\alpha_k,3|\hat H_e|\bm S_k,\alpha_k,3\rangle_a
\end{eqnarray}
and thus
\begin{eqnarray}
\lambda_{\bm S_k,\alpha_k,3}&=&E_{\bm S_k,\alpha_k,3}-\sum_{k}\lambda_{\bm S_k,1,2}\cdot{}_b\langle\bm S_k,\alpha_k,3|\hat V_{k,\alpha_k}^{\text{2-bdy}}|\bm S_k,\alpha_k,3\rangle_b\nonumber\\
&=&E_{\bm S_k,\alpha_k,3}-\sum_{k}E_{\bm S_k,1,2}\cdot{}_b\langle\bm S_k,\alpha_k,3|\hat V_{k,\alpha_k}^{\text{2-bdy}}|\bm S_k,\alpha_k,3\rangle_b
\end{eqnarray}
The computation of $\lambda_{\bm S_k,\alpha_k,n}$ with $n>3$ can be carried out similarly. The general recursive relation is given as follows:
\begin{eqnarray}
\lambda_{\bm S_k,\alpha_k,n}&=&E_{\bm S_k,\alpha_k,n}-\sum_{m=2}^{n-1}\sum_{k,\alpha_k}\lambda_{\bm S_k,\alpha_k,m}\cdot{}_b\langle\bm S_k,\alpha_k,n|\hat V_{k,\alpha_k}^{\text{m-bdy}}|\bm S_k,\alpha_k,n\rangle_b\label{recursion}
\end{eqnarray}
In principle, all terms in Eq.(\ref{recursion}) can be computed analytically in the thermodynamic limit, though we have not found a way to give the analytical expressions in a concise form. If we take $\hat H_e=\hat V^{\text{2-bdy}}_1$, which is the two-body Haldane pseudopotential, or the model Hamiltonian of the Laughlin $\nu=1/3$ phase, then we have in the thermodynamic limit the following expression:
\begin{eqnarray}
&&E_{\bm S_0,1,n}=\sum_{k_1,k_2=0}^{n-1}\frac{\left(k_1+k_2-1\right)!}{k_1!k_2!}\left(\frac{1}{2}\right)^{k_1+k_2}\left(k_1-k_2\right)^2\\
&&{}_b\langle\bm S_0,1,n|\hat V_{\bm S_0,1,m}|\bm S_0,1,n\rangle_b=\frac{n!}{m!\left(n-m\right)!}\qquad \text{for}\qquad m\le n
\end{eqnarray}

This allows us to compute the coefficients to the leading n-body pseudopotential in the effective Hamiltonian of Eq.(\ref{hb}) as follows:
\begin{eqnarray}
\lambda_{\bm S_0,1,n}=\sum_{k_1,k_2=0}^{n-1}\frac{\left(k_1+k_2-1\right)!}{k_1!k_2!}\left(\frac{1}{2}\right)^{k_1+k_2}\left(k_1-k_2\right)^2-\sum_{m=2}^{n-1}\lambda_{\bm S_0,1,m}\frac{n!}{m!\left(n-m\right)!}
\end{eqnarray}
\begin{figure}
\begin{center}
\includegraphics[width=\linewidth]{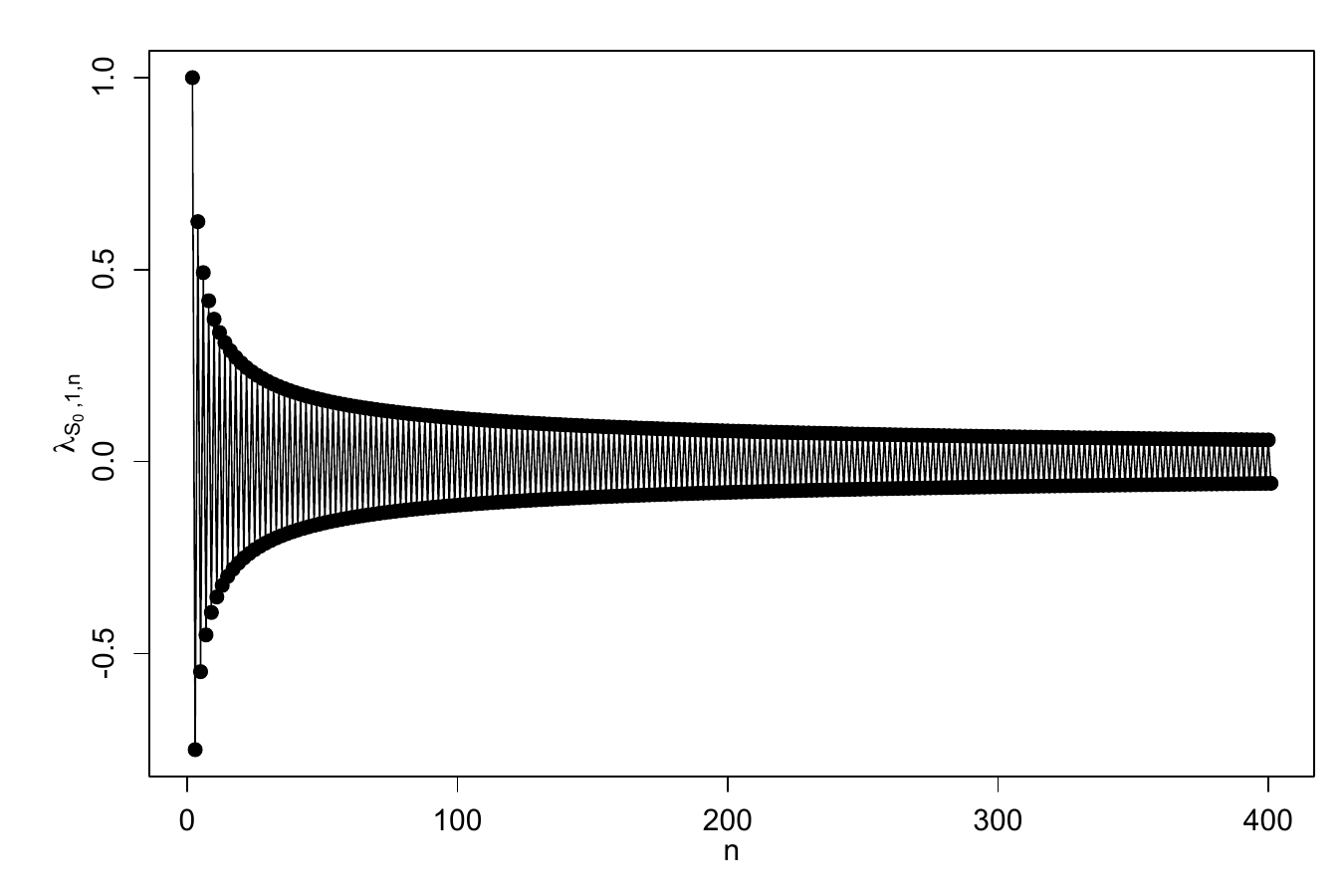}
\caption{The dependence of $\lambda_{\bm S_0,1,n}$ on $n$, when we bosonize fermionic fluxes in $\mathcal H_{[1,0]}$, with the bare interaction between electrons given by $\hat V_1^{\text{2bdy}}$. }
\label{fig3}
\end{center}
\end{figure}
All other terms not included in the summation above are actually zero, and the dependence of $\lambda_{\bm S_0,1,n}$ on $n$ is plotted in Fig.(\ref{fig3}). As we can see, $\lambda_{\bm S_0,1,n}$ alternates between positive and negative values, and will eventually decay to zero, albeit rather slowly. This is the manifestation of the long-range nature of the statistical interaction, and we conjecture this is true for other short-range interactions between electrons. For a system when the vacuum contains $28$ orbitals, we numerically calculate various different values of $\lambda_{\bm S_k,\alpha_k,n}$ with two types of bare electron interactions, as shown in Table.\ref{t2}. In general, the coefficients of the pseudopotentials decrease rather rapidly with increasing total angular momentum $k$ of the pseudopotentials. For Fig. 2 in the main text, the numerical computation is carried out with a truncated Hamiltonian with terms from the second column of Table.\ref{t2}, and the terms selected are: $\lambda_{\bm S_0,1,2},\lambda_{\bm S_0,1,3},\lambda_{\bm S_2,1,3},\lambda_{\bm S_3,1,3},\lambda_{\bm S_4,1,3},\lambda_{\bm S_0,1,4},\lambda_{\bm S_2,1,4},\lambda_{\bm S_3,1,4}$. Keeping these terms can already yield the ground state and the magnetoroton modes very similar to the corresponding fermionic FQH phase.
\begin{table}[h!]
\centering
\begin{tabular}{|c|c|c|c|c|} 
 \hline
 \multicolumn{1}{|c}{}&\multicolumn{1}{|c}{} & \multicolumn{1}{c}{$|\psi\rangle_{[1,0]}$ as the vacuum} & \multicolumn{1}{c|}{}& \multicolumn{1}{c|}{$|\psi\rangle_{[\frac{1}{3},-2]}$ as the vacuum}  \\
 \hline
  & $\hat H_e=\hat V^{\text{2bdy}}_1$ &$\hat H_e=\hat V^{\text{2bdy}}_3$ & $\hat H_e=\hat V^{\text{3bdy}}_3$&$\hat H_e=\hat V^{\text{2bdy}}_3$\\ 
 \hline
$\lambda_{\bm S_0,1,2}$ & $1$ & $0$ & $3.26$ &0.312\\ 
 \hline
 $\lambda_{\bm S_2,1,2}$ & $0$ & $1$ & $1.14$ &-0.331\\ 
 \hline
  $\lambda_{\bm S_4,1,2}$ & $0$ & $0$ & $0$ &0.159\\ 
 \hline
  $\lambda_{\bm S_6,1,2}$ & $0$ & $0$ & $0$ &-0.156\\ 
 \hline
   $\lambda_{\bm S_8,1,2}$ & $0$ & $0$ & $0$ &-0.0454\\ 
 \hline
  $\lambda_{\bm S_0,1,3}$ & $-0.737$ & $0.737$ & $-2.56$&-0.301\\ 
 \hline
   $\lambda_{\bm S_2,1,3}$ & $0.177$ & $-1.10$ & $-0.680$&0.270\\ 
 \hline
    $\lambda_{\bm S_3,1,3}$ & $-0.168$ & $-0.360$ & $-0.956$&0.0887\\ 
 \hline
    $\lambda_{\bm S_4,1,3}$ & $0.117$ & $-0.0185$ & $0.361$&-0.208\\ 
 \hline
     $\lambda_{\bm S_5,1,3}$ & $-0.0713$ & $0.239$ & $0.0403$&-0.140\\ 
 \hline
      $\lambda_{\bm S_6,1,3}$ & $0.0397$ & $-0.191$ & $-0.386$&0.00684\\ 
 \hline
       $\lambda_{\bm S_6,2,3}$ & $0$ & $0$ & $0.297$&0.0412\\ 
 \hline
        $\lambda_{\bm S_7,1,3}$ & $-0.0207$ & $0.0598$ & $0.000709$&0.0487\\ 
 \hline
         $\lambda_{\bm S_0,1,4}$ & $0.611$ & $-1.21$ & $2.22$&0.312\\ 
 \hline
          $\lambda_{\bm S_2,1,4}$ & $-0.232$ & $1.32$ & $0.519$&-0.203\\ 
 \hline
          $\lambda_{\bm S_3,1,4}$ & $-0.156$ & $0.0871$ & $0.772$&-0.00885\\ 
 \hline
           $\lambda_{\bm S_4,1,4}$ & $-0.240$ & $0.678$ & $0.719$&0.183\\ 
 \hline
            $\lambda_{\bm S_4,2,4}$ & $0.171$ & $-0.171$ & $-0.450$&0.0242\\ 
 \hline
             $\lambda_{\bm S_5,1,4}$ & $0.0417$ & $-0.249$ & $-0.112$&0.144\\ 
 \hline
              $\lambda_{\bm S_0,1,5}$ & $-0.534$ & $1.14$ & $-2.00$&\\ 
 \hline
               $\lambda_{\bm S_2,1,5}$ & $0.252$ & $0.780$ & $-0.447$&\\ 
 \hline
               $\lambda_{\bm S_3,1,5}$ & $-0.158$ & $-0.121$ & $-0.0237$&\\ 
 \hline
\end{tabular}
\caption{Coefficients of the effective Hamiltonians for bosons, in $\mathcal H_{[1,0]}$ and $\mathcal H_{[\frac{1}{3},-2]}$. The bare interaction between electrons is given by $\hat H_e$. All values of the coefficients are computed for a system where $|\psi\rangle_{[1,0]}$ and $|\psi\rangle_{[\frac{1}{3},-2]}$ contains $28$ orbitals. For angular momentum sectors with $\bar\alpha_{k,N_{\text{qh}}}>1$, we choose the gauge that gives the basis that is diagonal in $\hat H_b$.}
\label{t2}
\end{table}

We would also like to comment that while all the numerical computations are performed on the spherical geometry in this work, the bosonization scheme can be carried out in any geometries, since it fundamentally depends on the conformal symmetry of the Hilbert space (i.e. the bulk-edge correspondence can be defined on compact geometries via the entanglement spectrum). On the spherical geometry, the bosonization of fermions or anyons is carried out with the two good quantum numbers: the total angular momentum and the total z-component of the angular momentum. On the disk geometry the analogous procedure can be carried out using the total and center of mass angular momentum, while on the torus we can use the two linear momenta. The calculation of the spectrum of the bosonic Hamiltonians from the dual description, however, can lead to results with small quantitative differences (for example on the torus as compared to the sphere), if a truncated Hamiltonian is used, as is the case in the main text. Such quantitative differences should vanish in the thermodynamic limit, when the rotational invariance of the torus geometry is restored.

\section{S3. The bulk-edge correspondence}

In this section, we give a more detailed exposition of the role of the bulk-edge correspondence in our ability to bosonize fermions and anyons in a two-dimensional manifold. We first introduce the terminology of the ``conformal Hilbert spaces", which include the familiar null spaces of model Hamiltonians of the FQH phases (e.g. the pseudopotential Hamiltonians). As explained in the main text, a more general way of constructing the conformal Hilbert spaces is to use LEC\cite{yang3_sup}, and all such Hilbert spaces $\mathcal H_{\text{cfm}}$  satisfy the following properties in the thermodynamic limit (e.g. the number of electrons $N_e\rightarrow\infty$):
\begin{enumerate}
\item There is one highest density state in $\mathcal H_{\text{cfm}}$, denoted as $|\psi\rangle_0$;
\item The entanglement spectrum of $|\psi\rangle_0$ has a one-to-one correspondence to all other states in $\mathcal H_{\text{cfm}}$;
\item We can use conformal generators to define a set of highest weight states in $\mathcal H_{\text{cfm}}$, and $|\psi\rangle_0$ is one of the highest weight states;
\item All other states in $\mathcal H_{\text{cfm}}$ can be generated by conformal generators acting on the highest weight states.
\end{enumerate}
In the context of FQH, $|\psi\rangle_0$ corresponds to the ground state of a specific FQH phase, while all other states in $\mathcal H_{\text{cfm}}$ are the quasihole states. The statements above apply to the $\mathcal H_{\text{cfm}}$ with a fixed $N_e$, and are valid in the limit of large $N_e$. There can also be constraints on the allowed values of $N_e$. For example, $N_e$ is even for $\mathcal H_{\text{cfm}}$ corresponding to the Moore-Read phase, or the null space of the leading three-body pseudopotential $\hat V_3^{\text{3bdy}}$. We will also define the conformal generators properly later on.

The quasihole states are obtained from flux insertion to $|\psi\rangle_0$, so they are all less dense than $|\psi\rangle_0$. Let us now consider the geometry with an open boundary (e.g. the disk geometry). Due to the non-zero Hall conductivity, an insertion of the magnetic flux at a location $x_0$ within the ground state quantum fluid pushes electrons away from $x_0$, leading to a density fluctuation at the boundary/edge, while at the same time leaving behind a quasihole in the bulk. Thus every flux insertion can either be treated as a quasihole excitation in the bulk, or a density modulation at the edge. There is no ambiguity even if $x_0$ is deep in the bulk and far away from the edge: it is still an edge excitation, but with a very large momentum. For low-energy theory at the edge (as a one-dimensional system) in the long wavelength limit, only flux insertions near the edge are involved. However, the bulk-edge correspondence persists even for high energy excitations with large momenta.

It is thus straightforward to see that quasihole excitations on the disk geometry can be bosonized, since they can be understood as the edge excitations of the chiral Luttinger liquid of the one-dimensional boundary. Given that a conformal Hilbert space on the disk can be mapped to a conformal Hilbert space on the spherical geometry with the same model Hamiltonian or LEC, we can in principle also bosonize quasiholes on compact geometries like the sphere with no boundary. On such geometries, the quasihole excitations can only be considered as bulk excitations on the two-dimensional manifold. 

We will now show this explicitly by first noting that the counting of the quasihole states in different angular momentum sectors on the sphere can be mapped to the Virasoro counting of the edge excitations. Let a quasihole state on the sphere be denoted as $|\bm s_k,\bm S_k,\alpha_k\rangle_s$, where as before $\bm s_k,\bm S_k$ are the quantum numbers of $\hat L_z, \hat L^2$, with $\alpha_k$ denoting the degeneracy of the states with the same quantum numbers. On the disk, without loss of generality we use the symmetric gauge in the LLL. A many-body wavefunction is given by:
\begin{eqnarray}\label{states}
\psi_k\left(z_1,z_2\cdots z_{N_e}\right)=\phi_k\left(z_1,z_2\cdots z_{N_e}\right)e^{-\frac{1}{4}\sum_iz_iz_i^*}
\end{eqnarray}
where $z_i=x_i+iy$ are the holomorphic variables, and the subscript $i$ is the electron index. The polynomial $\phi_k\left(z_1,z_2\cdots z_{N_e}\right)$ is a linear combination of the monomial basis. We can thus write $\phi_k=\sum_\lambda c_{k\lambda}m_\lambda$, where $m_\lambda=\text{Asy}\left(z_1^{n_{1\lambda}}z_2^{n_{2\lambda}}\cdots z_{N_e}^{n_{N_e\lambda}}\right)$. The antisymmetrisation $\text{Asy}$ is over the electron indices. In addition, we assume rotational invariance on the disk geometry. If we define $Z=\sum_iz_i/N_e$, the center of mass angular momentum operator $\hat M\sim\partial_Z$ and the total angular momentum operator $\hat m=\sum_i\partial_{z_i}$ give two good quantum numbers of the disk wavefunctions. We can thus denote the quasihole states with $|m_k,M_k,\beta_k\rangle_d$, and $m_k=\sum_in_{i\lambda}$, which is independent of $\lambda$. The Gaussian factor in Eq.(\ref{states}) is not important, so we can just focus on $\phi_k$.

Every highest weight state on the sphere, $|\bm S_k,\bm S_k,\alpha_k\rangle_s$, is mapped to a state on the disk with no center of mass angular momentum: $|m_k,0,\beta_k\rangle_d$. In particular, let $|\bm S_k,\bm S_k,\alpha_k\rangle_s$ be the state containing $N_e$ electrons, $N_o$ orbitals and $N_{\text{qh}}$ quasiholes, with $\bm S_k=N_{\text{qh}}N_e/2-k$ and $\alpha_k=1,2,\cdots \bar\alpha_{k,N_{\text{qh}}}$. The values of $\bar\alpha_{k,N_{\text{qh}}}$ for different values of $k$ are listed in Table. \ref{t1}. It corresponds to the state on the disk $|m_0+k,0,\beta_k\rangle_d$ with the same number of electrons and orbitals, and $\beta_k=\alpha_k$, where $m_0=N_e\left(N_o-1\right)/2$ is the total angular momentum of the ground state. Let $\Delta m$ be the difference between the total angular momentum of the quasihole state and that of the ground state, and $p_{\Delta m}$ be the counting of the number of the quasihole states with total angular momentum $m_0+\Delta m$, we then have the following relationship for $\Delta m>0$:
\begin{eqnarray}
p_{\Delta m}=\lim_{N_{\text{qh}}\rightarrow\infty}\sum_{i=1}^{\Delta m}\bar\alpha_{i,N_{\text{qh}}}
\end{eqnarray}
We are interested in the system in the thermodynamic limit, where in the disk geometry we have a droplet of quantum fluid containing $N_e$ electrons embedded in an infinite plane (thus $N_{\text{qh}}\rightarrow\infty$), and taking $N_e\rightarrow\infty$. Using the partition theory\cite{partition_sup} we can show $\bar\alpha_{i,N_{\text{qh}}\rightarrow\infty}=\mathcal P\left(i\right)-\mathcal P\left(i-1\right)$, where $\mathcal P\left(i\right)$ is the integer partition of $i$, or the number of different ways of writing $i$ as a sum of positive integers. This immediately gives us $p_{\Delta m}=\mathcal P\left(\Delta m\right)$, which is the Virasoro counting of the edge excitation corresponding to the chiral Luttinger liquid.

Having established that the counting of the bulk quasihole excitations reproduces the Virasoro counting of the edge excitations, we have shown that the bulk quasiholes can in principle be mapped to the bosonic density modes of the edge excitations of the same FQH phase. In the main text, the unitary transformation for such mapping is explicitly constructed. Note that the universality of such mapping is based on the assumption that the Hilbert space of the edge excitations, or the null space of the corresponding model Hamiltonian, is a good representation of the Virasoro algebra. This has been well established from the effective description exploiting the conformal symmetry of the $1+1D$ chiral edge system\cite{read2_sup}. In principle, since we are constructing the bosonization formalism microscopically, this conformal symmetry should also be established microscopically. This is a rather non-trivial process that has been detailed in Ref.\cite{boyang_sup}. Here we give a brief outline of the main ideas. 

For many-body wavefunctions on the disk geometry, we can define $c_i^\dagger$ as the electron creation operator in the single particle orbital indexed by $i$(i.e. $z^i$), satisfying the anticommutation relations $\{c_i,c_j^\dagger\}=\delta_{ij},\{c_i,c_j\}=\{c_i^\dagger,c_j^\dagger\}=0$. In the second quantized form, monomials in $\phi_k$ are given by:
\begin{eqnarray}
 m_\lambda=\text{Asy}\left(z_1^{n_{1\lambda}}z_2^{n_{2\lambda}}\cdots z_{N_e}^{n_{N_e\lambda}}\right)\sim c_{n_{1\lambda}}^\dagger c_{n_{2\lambda}}^\dagger\cdots c_{n_{N_e\lambda}}^\dagger|\text{vac}\rangle\qquad
 \end{eqnarray}
We now need to properly define the Virasoro generators acting on $\phi_k$ satisfying the Virasoro algebra given as follows\cite{read2_sup,textbook_sup}:
\begin{eqnarray}\label{virasoro}
 [\hat {\mathcal L}_n,\hat {\mathcal L}_m]=\left(n-m\right)\hat {\mathcal L}_{m+n}+\frac{c}{12}n\left(n^2-1\right)\delta_{m+n,0}
 \end{eqnarray}
Using the second quantised operators, for $n\ge 0$ we have the following construction:
\begin{eqnarray}
\hat L_{-n}=\sum_{k=0}^\infty f_{k+n,k}\cdot k\hat c^\dagger_{k+n}\hat c_k\label{vv1},\quad\hat L_{n}=\sum_{k=0}^\infty f_{k,k+n}\cdot \left(k+n\right)\hat c^\dagger_{k}\hat c_{k+n}\label{vv2}
\end{eqnarray}
The function $f_{k_1,k_2}$ comes from the single particle state normalisation, and on the disk geometry it is given by $f_{k_1,k_2}=\sqrt{k_1!/k_2!}$. It is easy to check that the Virasoro algebra is satisfied between $\hat L_m,\hat L_n$ if $n\cdot m\ge 0$. The commutation relation between the positive and negative modes is given as follows:
\begin{eqnarray}
&&[\hat L_{m},\hat L_{-n}]=\left(n-m\right)\hat L_{m+n}+\hat C_{m,n}\label{vv3},\quad\hat C_{m,n}=\begin{cases}
\sum\limits_{k=0}^{m-1}f_{k,k+\Delta}\cdot\left(k-m\right)\left(k+\Delta\right)\hat c_k^\dagger\hat c_{k+\Delta}&n\ge m\qquad\\
\sum\limits_{k=0}^{n-1}f_{k+\Delta,k}\cdot k\left(k-n\right)c^\dagger_{k+\Delta}\hat c_k&n\le m\qquad
\end{cases}
\end{eqnarray}
with $m,n\ge 0, \Delta=|m-n|$. The central charge emerges from the additional term $\hat C_{m,n}$ in the thermodynamic limit, after projection into the the null space spanned by the ground state and the quasiholes. The projection is necessary because $\hat L_n$ acting on the quasihole and ground states generally will create gapped excitations beyond the null space. The conformal symmetry is also only strictly obeyed in the thermodynamic limit; this is also well know because the counting of the edge excitations does not fully obey the Viraroso counting for finite systems\cite{hermanns_sup}. 

Thus starting with the Laughlin ground state $|\phi_L\rangle$ with $\langle z_1,z_2,\cdots z_{N_e}|\phi_L\rangle\sim\prod_{i<j}\left(z_i-z_j\right)^3$, the entire Hilbert space spanned by the Laughlin quasiholes can be obtained by linearly independent states generated as follows:
\begin{eqnarray}
|\phi_{qh},\Delta m\rangle\sim \hat L_{-n_1}\hat L_{-n_2}\cdots\hat L_{-n_k}|\phi_L\rangle
\end{eqnarray}
where $n_1+n_2\cdots +n_k=\Delta m$, which is an integer partition of $\Delta m$. Thus at the microscopic level, we have shown the conformal symmetry of the quasihole states on the disk, which is required for the bosonization of the edge excitations. The mapping of the quasiholes on the disk and spherical geometry (or other compact geometries), also at the microscopic level, illustrates the interplay between 2D bosonization, bulk-edge correspondence and conformal symmetry in condensed matter systems.


\begin{thebibliography}{99}
\bibitem{Myrheim} J.M. Leinaas and J. Myrheim, Nuovo Cim. B 37, 1 (1977).
\bibitem{wilczek1} F. Wilczek, Phys. Rev. Lett. {\bf 49}, 957 (1982).
\bibitem{wilczek2}D. Arovas, J.R. Schrieffer, and F. Wilczek, Phys. Rev. Lett. {\bf 53}, 722 (1984).
\bibitem{mr} G. Moore and N. Read, Nucl. Phys. B {\bf 360}, 362 (1991).
\bibitem{kitaev} A. Kitaev, Ann. Phys. {\bf 303}, 2 (2003).
\bibitem{nayak}  C. Nayak, S.H. Simon, A. Stern, M. Freedman, and S. Das Sarma, Rev. Mod. Phys. {\bf 80}, 1083 (2008).
\bibitem{prange} R. Prange and S. Girvin, The Quantum Hall Effect, Graduate Texts in Contemporary Physics (Springer-Verlag, Berlin, 1987).
\bibitem{laughlin} R.B. Laughlin, Phys. Rev. Lett. {\bf 50} 1395 (1983).
\bibitem{halperin}B.I. Halperin, Phys. Rev. Lett. {\bf 52}, 1583 (1984).
\bibitem{stern} A. Stern, Ann. Phys. {\bf 323}, 204 (2008).
\bibitem{clarke} D.J. Clarke, J. Alicea, and K. Shtengel, Nat. Comm. {\bf 4}, 10.1038/ncomms2340 (2013).
\bibitem{banerjee}  M. Banerjee, M. Heiblum, V. Umansky, D.E. Feldman,Y. Oreg, and A. Stern, Nature {\bf 559}, 205 (2018).
\bibitem{manfra} J. Nakamura, S. Liang, G.C. Gardner and M.J. Manfra, Nat. Phys. {\bf 16}, 931 (2020).
\bibitem{ajit} A.C. Balram, J. Jain, and M. Barkeshli, Phys. Rev. Res. {\bf 2}, 013349 (2020).
\bibitem{yang}Ha Quang Trung and Bo Yang, arXiv: 2009.14214.
\bibitem{haldane1} F.D.M. Haldane, Phys. Rev. Lett. {\bf 67}, 937 (1991).
\bibitem{rr} N. Read and E. Rezayi, Phys. Rev. B, {\bf 59} 8084 (1999).
\bibitem{zee} D.P. Arovas, R. Schrieffer, F. Wilczek and A. Zee, Nucl. Phys. B. {\bf 251}, 117 (1985).
\bibitem{ragnu} Jing-Yuan Chen, Jun Ho Son, Chao Wang and S. Raghu, Phys. Rev. Lett. {\bf 120}, 016602 (2018).
\bibitem{jensen} K. Jensen, J. High Energy Phys. {\bf 2018}, 31 (2018).
\bibitem{mulligan} A. Hui, Eun-Ah Kim and M. Mulligan, Phys. Rev. B. {\bf 99}, 125135 (2019).
\bibitem{son} W-H. Hsiao and D.T. Son, Phys. Rev. B. {\bf 100}, 235150 (2019).
\bibitem{fradkin} Y. Ferreiros and E. Fradkin, Ann. Phys. {\bf 399}, 1 (2018).
\bibitem{seidel} T. Mazaheri, G. Ortiz, Z. Nussinov and A. Seidel, Phys. Rev. B. {\bf 91}, 085115 (2015).
\bibitem{haldane} H. Li and F.D.M. Haldane, Phys. Rev. Lett. {\bf 101}, 010504 (2008).
\bibitem{yan} B. Yan, R.R. Biswas and C.H. Greene, Phys. Rev. B. {\bf 99}, 035153 (2019).
\bibitem{anushya}A. Chandran, M. Hermanns, N. Regnault, and B. Andrei Bernevig, Phys. Rev. B. {\bf 84}, 205136 (2011).
\bibitem{chang} A.M. Chang, Rev. Mod. Phys. {\bf 75}, 1449 (2003).
\bibitem{wen} X.G. Wen, Phys. Rev. B. {\bf 41}, 12838 (1990).
\bibitem{cooper} S.H. Simon, E.H. Rezayi and N.R. Cooper, Phys. Rev. B. {\bf 75}, 195306 (2007).
\bibitem{jackson} T.S. Jackson, N. Read and S.H. Simon, Phys. Rev. B. {\bf 88}, 075313 (2013).
\bibitem{santos} L.H. Santos, Phys. Rev. Research, {\bf 2}, 013232 (2020).
\bibitem{kapustin} Y-A Chen and A. Kapustin, Phys. Rev. B. {\bf 100}, 245127 (2019).
\bibitem{sphere} F.D.M. Haldane, Phys. Rev. Lett. {\bf 51}, 605 (1983).
\bibitem{geometry}Generalisation to other geometries is straightforward and is discussed in \cite{supp}.
\bibitem{yang1} Bo Yang, Z. Papic, E.H. Rezayi, R.N. Bhatt and F.D.M Haldane, Phys. Rev. B. 85, 165318 (2012).
\bibitem{yang2} Bo Yang, Zixiang Hu, Ching Hua Lee and Zlatko Papic, Phys. Rev. Lett. 118, 146403 (2017).
\bibitem{simon} S.H. Simon, E.H. Rezayi, and N.R. Cooper, Phys. Rev. B {\bf 75}, 075318 (2007).
\bibitem{yang3} Bo Yang, Phys. Rev. B. {\bf 100}, 241302(R) (2019).
\bibitem{wenzee}X.G. Wen and A. Zee, Phys. Rev. B. {\bf 46}, 2290 (1992).
\bibitem{readshift} N. Read and E.H. Rezayi, Phys. Rev. B. {\bf 84}, 085316 (2011).
\bibitem{bernevig} B.A. Bernevig and F.D.M. Haldane, Phys. Rev. Lett. 100, 246802 (2008).
\bibitem{jain} J.K. Jain, Composite Fermions (Cambridge University Press, New York, US, 2007).
\bibitem{ajit2} A.C. Balram, A. Wojs and J.K. Jain, Phys. Rev. B. {\bf 88}, 205312 (2013).
\bibitem{toke} C. Toke and J.K. Jain, Phys. Rev. B. {\bf 80}, 205301 (2009).
\bibitem{ywz} Bo Yang, Y-H. Wu and Z. Papic, Phys. Rev. B. {\bf 100}, 245303 (2019).
\bibitem{read} N. Read, Phys. Rev. B. {\bf 79}, 245304 (2009).
\bibitem{preskill} A. Kitaev and J. Preskill, Phys. Rev. Lett. {\bf 96}, 110404 (2006).
\bibitem{levin} M. Levin and X.-G. Wen, Phys. Rev. Lett. {\bf 96}, 110405 (2006).
\bibitem{senthil} T. Senthil and M. Levin, Phys. Rev. Lett. {\bf 110}, 046801 (2013).
\bibitem{vish} Yin-Chen He, Fabian Grusdt, Adam Kaufman, Markus Greiner, and Ashvin Vishwanath, Phys. Rev. B {\bf 96}, 201103(R) (2017).
\bibitem{yinghai} Ying-Hai Wu and J.K. Jain, Phys. Rev. B {\bf 87}, 245123 (2013).
\bibitem{private} A.C. Balram, private communications.
\bibitem{supp} See Supplemental Material at [url] for detailed calculation and analysis, which includes Refs. \cite{partitionsupp,boyangsupp,textbooksupp,hermannssupp}.
\bibitem{partitionsupp} Andrews, George E. (1976). The Theory of Partitions. (Cambridge University Press).
\bibitem{boyangsupp} Bo Yang, Phys. Rev. B. {\bf 103}, 115102 (2021).
\bibitem{textbooksupp} F. Di Francesco, P. Mathieu, and D. Sénéchal, Conformal Field Theory (Springer, Berlin, 1997).
\bibitem{hermannssupp} M. Hermanns, A. Chandran, N. Regnault, and B. Andrei Bernevig, Phys. Rev. B. {\bf 84}, 121309(R) (2011).
\bibitem{footnote} Note that we can still bosonize the MR state within the Hilbert space of a single LL (which is abelian). For example the MR ground state with $N_e$ electrons is a quantum fluid with $N_e-2$ quasiholes (when the vacuum is the fully filled LL), which can be bosonized in a completely analogous manner as the Laughlin states.
\bibitem{sarma} M. Storni, R.H. Morf and S.Das Sarma, Phys. Rev. B. {\bf 104}, 076803 (2010).
\bibitem{hlr} B.I. Halperin, P.A. Lee and N. Read, Phys. Rev. B. {\bf 47}, 7312 (1993).
\bibitem{sondirac}  D.T. Son, Phys. Rev. X {\bf 5}, 031027 (2015).
\bibitem{dx} D.X. Nguyen and D.T. Son, arXiv: 2105.02092.
\bibitem{shankar} G. Murphy and R. Shankar, Rev. Mod. Phys. {\bf 75}, 1101 (2003).
\bibitem{ajit3} A.C. Balram, J.K. Jain and M. Barkeshili, Phys. Rev. Research {\bf 2}, 013349 (2020).
\end{thebibliography}

\begin{thebibliography}{99}
\bibitem{yang3_sup} Bo Yang, Phys. Rev. B. {\bf 100}, 241302(R) (2019).
\bibitem{partition_sup} Andrews, George E. (1976). The Theory of Partitions. (Cambridge University Press).
\bibitem{read2_sup} N. Read, Phys. Rev. B. {\bf 79}, 245304 (2009).
\bibitem{boyang_sup} Bo Yang, Phys. Rev. B. {\bf 103}, 115102 (2021).
\bibitem{textbook_sup} F. Di Francesco, P. Mathieu, and D. Sénéchal, Conformal Field Theory (Springer, Berlin, 1997).
\bibitem{hermanns_sup} M. Hermanns, A. Chandran, N. Regnault, and B. Andrei Bernevig, Phys. Rev. B. {\bf 84}, 121309(R) (2011).
\end{thebibliography}
\end{document}